\documentclass[18pt]{article}

\usepackage{graphicx}

\usepackage{here}

\usepackage{latexsym}

\usepackage{amsbsy}

\usepackage{amsmath,amsthm,amssymb, amsfonts}

\usepackage[a4paper,left=27mm,right=25mm,top=35mm,bottom=40mm]{geometry}

\makeatletter
\renewcommand\section{\@startsection{section}{1}{\z@}%
{-3.25ex\@plus -1ex \@minus -.2ex}{1.5ex \@plus .2ex}%
{\normalfont\large\bfseries}}
\renewcommand\subsection{\@startsection{subsection}{2}{\z@}%
{-3.25ex\@plus -1ex \@minus -.2ex}%
{1.5ex \@plus .2ex}%
{\normalfont\normalsize\bfseries}}
\renewcommand\subsubsection{\@startsection{subsubsection}{3}{\z@}%
{-3.25ex\@plus -1ex \@minus -.2ex}%
{1.5ex \@plus .2ex}%
{\normalfont\normalsize\bfseries}}
\renewcommand\paragraph{\@startsection{paragraph}{4}{\parindent}%
{3.25ex \@plus1ex \@minus .2ex}%
{-1em}%
{\normalfont\normalsize\bfseries}}
\makeatother

\newcommand{\pa}{\mathrm{pa}}
\newcommand{\nei}{\mathrm{ne}}
\newcommand{\spo}{\mathrm{sp}}
\newcommand{\an}{\mathrm{an}}

\newcommand{\dse}{\,\mbox{$\perp$}\,}

\newcommand{\cd}{\,|\,}

\newcommand{\nl}{\vspace{3mm}\\}

\newcommand{\n}[0]{\hspace*{.35em}}

\newcommand{\nn}[0]{\hspace*{.7em}}

\newcommand{\ful}{\mbox{$\, \frac{ \nn \nn \;}{ \nn \nn
}$}}

\newcommand{\full}{\mbox{$\, \frac{ \nn \;}{ \nn
}$}}

\newcommand{\fla}{\mbox{$\hspace{.05em} \prec
\!\!\!\!\!\frac{\nn \nn}{\nn}$}}

\newcommand{\fra}{\mbox{$\hspace{.05em} \frac{\nn
\nn}{\nn
}\!\!\!\!\! \succ \! \hspace{.25ex}$}}

\newcommand{\fraa}{\mbox{$\hspace{.05em} \frac{
\nn}{\nn
}\!\!\!\!\! \succ \! \hspace{.25ex}$}}

\newcommand{\arc}{\mbox{$\hspace{.06em} \prec
\!\!\!\!\!\frac{\nn \nn}{\nn}
\!\!\!\!\!
\succ\! \hspace{.25ex}$}}

\newcommand{\arcc}{\mbox{$\hspace{.06em} \prec
\!\!\!\!\!\frac{\nn \n}{\nn}
\!\!\!\!\!
\succ\! \hspace{.25ex}$}}

\newtheorem{prop}{Proposition}
\newtheorem{coro}{Corollary}

\newtheorem{lemma}{Lemma}

\newtheorem{alg}{Algorithm}
\newtheorem{theorem}{Theorem}

\usepackage{here}
\usepackage{setspace}
\usepackage{natbib}
\usepackage{array}
\usepackage{arydshln}
\usepackage{graphicx} \graphicspath{{pics/}}
\usepackage{natbib}
\usepackage[all]{xy}

\begin{document}
\markright{
}
\markboth{\hfill{\footnotesize\rm Kayvan Sadeghi}\hfill}
{\hfill {\footnotesize\rm  Markov Equivalences for Subclasses of Loopless Mixed Graphs} \hfill}
\renewcommand{\thefootnote}{}
$\ $\par
\fontsize{12}{14pt plus.8pt minus .6pt}\selectfont
\vspace{0.8pc}
\noindent{\Large \bf Markov Equivalences for Subclasses of Loopless Mixed Graphs \\[6mm]}
\noindent{\large KAYVAN SADEGHI\\[2mm]}
{\it Department of Statistics, University of Oxford\\[4mm] }

\noindent{\bf ABSTRACT}: In this paper we discuss four problems regarding Markov equivalences for subclasses of loopless mixed graphs. We classify these four problems as finding conditions for internal Markov equivalence, which is Markov equivalence within a subclass, for external Markov equivalence, which is Markov equivalence between subclasses, for representational Markov equivalence, which is the possibility of a graph from a subclass being Markov equivalent to a graph from another subclass, and finding algorithms to generate a graph from a certain subclass that is Markov equivalent to a given graph. We particularly focus on the class of maximal ancestral graphs and its subclasses, namely regression graphs, bidirected graphs, undirected graphs, and directed acyclic graphs, and present novel results for representational Markov equivalence and algorithms.
\\[-1mm]
\section{Introduction}
\paragraph{Introduction and motivation.} In graphical Markov models several classes of graphs have been used in recent years. A common feature of all these graphs is that their nodes correspond to random variables, and they represent conditional independence
statements of the node set by specific interpretations of missing edges.

These graphs contain up to three different types of edges. \citet{sad11} gathered most classes of graphs defined in the literature under a unifying class of loopless mixed graphs (LMGs). These contain \emph{Summary graphs} (SGs) \citep{wer11}, \emph{(maximal) ancestral graphs} (MAGs) \citep{ric02}, \emph{acyclic directed mixed graphs} (ADMGs) \citep{spi97}, regression chain graphs (RCGs) \citep{cox93,wer04,wers11}, \emph{undirected} or \emph{concentration graphs} (UGs) \citep{dar80,lau96}, \emph{bidirected} or \emph{covariance graphs} (BGs) \citep{cox93,wer98}, and directed acyclic graphs (DAGs) \citep{kii84,lau96}.

For the above graphs, in general, two graphs of different types or even two graphs of the same type may induce the same independencies. Such graphs are said to be \emph{Markov equivalent}. It is important to explore the similar characteristics of Markov equivalent graphs, and to find the ways of generating graphs of a certain type with the same independence structure from a given graph.

\paragraph{Some questions for Markov equivalences.} There are four main questions
 regarding Markov equivalence for different types of graphs:
\begin{description}
	\item[1) Internal Markov equivalence:]The first natural question that arises in this context is regarding when two graphs of the same type are Markov equivalent. This question may be answered for DAGs, MAGs, or other subclasses of LMGs.
	\item[2) External Markov equivalence:]In addition to Markov equivalence for graphs of the same type, one can discuss Markov equivalence between two graphs of different types.
	\item[3) Representational Markov equivalence:]Before checking external Markov equivalence, however, it is essential to check whether and under what conditions a graph of a certain type can be Markov equivalent to a graph of another type.
	\item[4) Algorithms:] One can also present some algorithms to generate a graph of a certain type that is Markov equivalent to a given graph of a different type.
	\end{description}
In this paper we gather and simplify the existing results in the literature for internal and external Markov equivalences, and give novel results for representational Markov equivalence and algorithms.
\paragraph{Some earlier results on Markov equivalence for graphs.}
Results concerning Markov equivalence for different classes of graphs have been obtained independently in the statistical literature on specifying types of multivariate statistical models, and in
the computer science literature on deciding on special properties of a given graph or on
designing fast algorithms for transforming graphs. In the literature on graphical Markov models two of the early results concerning Markov equivalence for DAGs and chain graphs were respectively given in \citet{ver90} and \citet{fry90}. Two of the later results by \citet{zha04} and \citet{ali09} respectively provided theoretically neat and computationally fast conditions for Markov equivalence for maximal ancestral graphs.

Besides these, \citet{pea94} provided conditions for Markov equivalence for bidirected graphs and DAGs. \citet{spir97} gave some conditions for Markov equivalence for maximal ancestral graphs, in which the polynomial computational complexity claim was wrong.

Efficient algorithms for deciding whether a UG can be oriented into a DAG became available in the computer science literature under the name of perfect elimination orientations; see \citet{tar84}, whose algorithm can be run in $O(|V|+|E|)$. Another such linear algorithm can be found in \citet{bla92}.  An algorithm for generating a Markov equivalent DAG from a bidirected graph is the special case of the algorithm given in \citet{zha04}.
\paragraph{Structure of the paper.} In the next section we define the unifying class of LMGs, and provide some basic graph theoretical definitions needed for our results.

In Section 3 we present the subclasses of LMGs, and we formally define the subclasses of interest in this paper. We also define a so-called separation criterion, called $m$-separation, to provide an interpretation of independencies for the graphs.

In Section 4 we formally define Markov equivalence, define maximality and explain its importance for Markov equivalences, and motivate why we consider Markov equivalence for the class of MAGs.

In Section 5 we gather the conditions existing in the literature for internal Markov equivalence for the class of MAGs and its subclasses, and give conditions for their external Markov equivalence.

In Section 6  we discuss the conditions for representational Markov equivalence for MAGs and its subclasses to a specific subclass, and we also provide algorithms to generate a Markov equivalent graph of a specific type to a given graph of another type when the conditions for representational Markov equivalence are satisfied. In each subsection we deal with different subclasses of MAGs: DAGs in Section 6.1, UGs and BGs in Section 6.2, and RCGs in Section 6.3.

In Section 7 we summarise the results, presented in the paper.
\section{Loopless mixed graphs}
\paragraph{Graphs.} A \emph{graph} $G$ is a triple consisting of a \emph{node} set or
\emph{vertex} set $V$, an \emph{edge} set $E$, and a relation that with
each edge associates two nodes (not necessarily distinct), called
its \emph{endpoints}. A \emph{loop} is an edge with
the same endpoints. When nodes $i$ and $j$ are the endpoints of an
edge, these are
\emph{adjacent} and we write $i\sim j$. We say the edge is \emph{between} its two
endpoints. We usually refer to a graph as an ordered
pair $G=(V,E)$. Graphs $G_1=(V_1,E_1)$ and $G_2=(V_2,E_2)$ are called \emph{equal} if $(V_1,E_1)=(V_2,E_2)$. In this case we write $G_1=G_2$.

Notice that the graphs that we use in this paper (and in general in the context of graphical models) are so-called \emph{labeled graphs}, i.e.\ every node is considered  a different object. Hence, for example, graph $i\ful j\ful k$ is not equal to $j\ful i\ful k$.
\paragraph{Definition of loopless mixed graph.} \citet{sad11} gathered most graphs in the literature of graphical models under the definition of \emph{loopless mixed
graph}, which is a graph that contains three types of edges denoted by
arrows, arcs (two-headed arrows), and lines (full lines) and does not contain any loops.

\paragraph{Basic definitions for LMGs.} We say that $i$ is a
\emph{neighbour} of $j$ if these are endpoints of a line, and $i$ is a parent of $j$ if there is an arrow from $i$ to $j$. We also define that $i$ is a \emph{spouse} of $j$ if these are endpoints of an arc. We use the notations $\nei(j)$, $\pa(j)$, and $\spo(j)$ for the set of all neighbours, parents, and spouses of $j$ respectively. In the cases of $i\fra j$ or $i\arc j$ we say that there is an arrowhead pointing to (at) $j$

A \emph{subgraph} of a graph $G_1$ is graph $G_2$ such that $V(G_2)\subseteq V(G_1)$ and $E(G_2)\subseteq E(G_1)$ and the assignment of endpoints to edges in $G_2$ is the same as in $G_1$.  An \emph{induced subgraph} by nodes $A\subseteq V$ is a subgraph that contains all and only nodes in $A$ and all edges between two nodes in $A$. A subgraph induced by edges $A\subseteq E$ is a subgraph that contains all and only edges in $A$ and all nodes that are endpoints of edges in $A$. We denote the subgraphs induced
by arrows, arcs, and lines in a graph $H$ by $H[\fraa]$, $H[\arcc]$, and $H[\full]$ respectively.

A \emph{walk} is a list $\langle v_0,e_1,v_1,\dots,e_k,v_k\rangle$ of nodes and edges such that for $1\leq i\leq k$, the edge $e_i$ has endpoints $v_{i-1}$ and $v_i$. A \emph{path} is a walk with no repeated node or edge. We denote a path by an ordered sequence of node sets. We say a path is \emph{between} the first and the last nodes of the list in $G$. We
call the first and the last nodes \emph{endpoints} of the path and all other nodes \emph{inner nodes}.

A \emph{cycle} in a graph G is a simple subgraph whose nodes can be placed around a circle so that two nodes are
adjacent if these appear consecutively along the circle.

A path (or cycle) is \emph{direction
preserving} if all its edges are arrows pointing to one direction. If a direction-preserving path is from a node $j$ to a node $i$ then $j$ is an \emph{ancestor} of $i$. We denote the set of ancestors of $i$ by $\an(i)$.

A graph is called \emph{directed} if it only contains arrows. A directed graph is \emph{acyclic} if it has no direction-preserving cycle.

A \emph{chord} is an edge between two non-adjacent nodes on the cycle. A cycle is \emph{chordless} if it has no chords. The notation $C_n$ is used for a chordless cycle with $n$ nodes. Notice that $C_n$ can contain different types of edges, so it represents a class of graphs rather than a single graph. We call a graph \emph{chordal} if it has no $C_n$, $n\geq 4$, as an induced subgraph. We also use the notation $P_n$ for a \emph{chordless} or \emph{minimal} path with $n$ nodes, i.e.\ a path that has no edge between two non-adjacent nodes on the path.

A \emph{V-configuration} is a path with three nodes and two edges. Notice that originally and in most texts, e.g.\ in \citet{kii84}, the endpoints of a V-configuration is by definition not adjacent. In this paper we call such V-configurations \emph{unshielded}.

In a mixed graph the inner node of three V-configurations $i\fra\,
t\fla\,j$, $i\arc\,t\fla\,j$, and $i\arc\,t\arc\,j$ is
a \emph{collider} and the inner node of all other V-configurations
is a \emph{non-collider} node on the V-configuration or more generally on a path of which the V-configuration is a subpath. We also call the V-configuration with collider or non-collider inner node a \emph{collider} or \emph{non-collider V-configuration} respectively. We may speak of a collider or non-collider node without
mentioning the V-configuration or the path when this is apparent from the context. In the case of DAGs the only collider V-configuration $i\fra\,
t\fla\,j$ is called a collision V-configuration.
\section{Subclasses of loopless mixed graphs and their independence interpretation}
\paragraph{Subclasses of LMGs} The following diagram, presented in \citet{sad11} illustrates the hierarchy regarding inclusions of subclasses of LMGs.
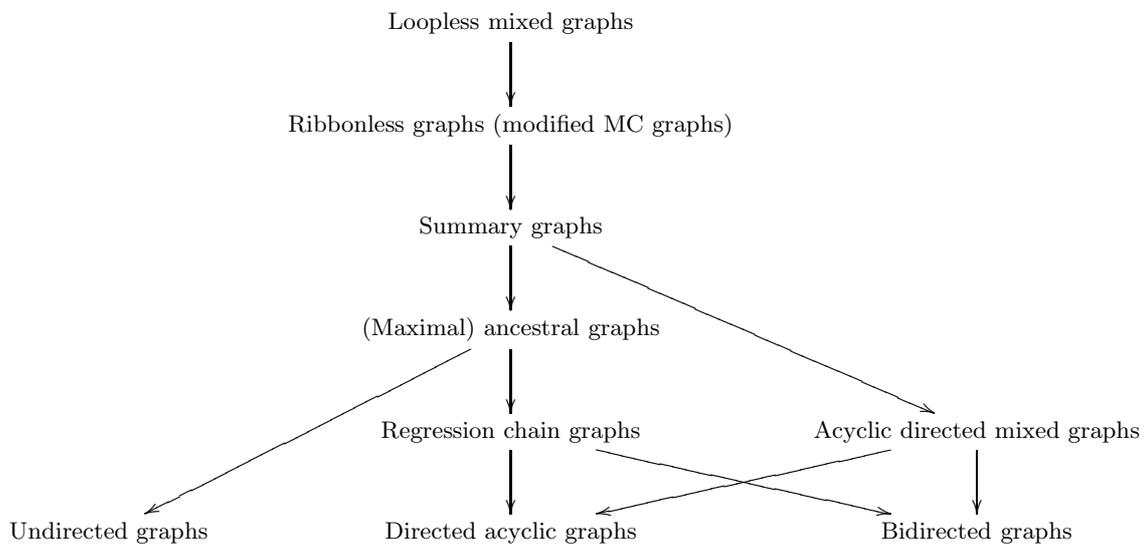
\begin{figure}[H]
            \centering\n
            \small{\xymatrix{&\txt{Loopless mixed graphs}\ar[d] \\ & \txt{Ribbonless graphs (modified MC graphs)}\ar[d] \\ &\txt{Summary graphs}\ar[d]\ar[rdd] \\ & \txt{(Maximal) ancestral graphs}\ar[d]\ar[ldd] \\  & \txt{Regression chain graphs}\ar[d]\ar[rd]
             &\txt{Acyclic directed mixed graphs}\ar[d]\ar[ld] \\ \txt{Undirected graphs} & \txt{Directed acyclic graphs} & \txt{Bidirected graphs}}}
            \caption[]{\small{The hierarchy of subclasses of LMGs.}}
        \label{fig:1110}
\end{figure}
The common feature of all these subclasses is that these use the same interpretation for independencies, known as the $m$-separation criterion. We will shortly introduce the $m$-separation criterion for MAGs and its subclasses.

\emph{Ribbonless graphs}, defined and studied in \citet{sad112} is a modification of MC graphs, defined by \citet{kos02}, to discard the line loops and to use the $m$-separation criterion.
\paragraph{Definition of ancestral graphs and regression chain graphs.} An \emph{ancestral graph} is a mixed graph that, for all nodes $i$, has (1) $i \notin \an(\pa(i)\cup\spo(i))$ and  (2) If $\nei(i)\neq \varnothing$ then $\pa(i)\cup\spo(i)=\varnothing$. This means that there is no arrowhead pointing to a line and there is no direction-preserving cycle, and there is no arc with one endpoint the ancestor of the other endpoint, in the graph.

A graph $G=(V,E)$ is a \emph{regression chain graph} if it contains at most the three types of edge, there is no arrowhead pointing to a line in graph, and it does not contain any \emph{arc-direction-preserving} cycle, i.e.\ a cycle that contains arcs and at least one arrow and whose arrows are all towards one direction. Thus in such graphs the subgraph induced by lines is so-called at the beginning of graph, and the subgraph induced by the arrows and arcs is characterised by having a node set that can be partitioned into numbered subsets forming so-called \emph{chains}, i.e.\
$V=\tau_1\cup\dots\cup\tau_T$ such that all edges between nodes in the same subset are arcs and all
edges between different subsets are arrows pointing from the set with the higher number to the one with the lower number. One can observe that in the subgraph induced by the arrows and arcs if we replace every $\tau_i$ by a node, we get a DAG.
\paragraph{The $m$-separation criterion}  Since, as we shall see, we are only interested in the subclasses of MAGs, we use the simplified version of $m$-separation criterion, defined in \citet{sad11}. This is identical to the original definition of $m$-separation; see \citet{ric02}.

Let $C$ be a subset of the node set $V$ of a MAG. A path is $m$-connecting given $C$ if all its
collider nodes are in $C\cup\an(C)$ and all its non-collider nodes are outside $C$.  For two other disjoint subsets of the node set $A$ and $B$, we say $A\dse_mB\cd C$ if there is no $m$-connecting path between $A$ and $B$ given $C$.

Notice that the $m$-separation criterion gives an interpretation of independencies on graphs, i.e.\ it induces an independence model.
\section{Basic concepts for Markov equivalence}
\paragraph{Definition of Markov equivalence.} Now we can formally define Markov equivalence. Two graphs $G_1=(V,E_1)$ and $G_2=(V,E_2)$ are \emph{Markov equivalent} if, for all  subsets $A$, $B$, and $C$ of $V$, $A\dse_m B\cd C$ in $G_1$ if and only if $A\dse_m B\cd C$ in $G_2$.
\paragraph{Maximality and Markov equivalence.}  A loopless mixed graph $G$ is called \emph{maximal} if by adding any edge to $G$ the independence model induced by the $m$-separation criterion changes. Alternatively, a graph $G=(V,E)$ is maximal if and only if, for every pair of non-adjacent nodes $i$ and $j$ of $V$, there exists a subset $C$ of $V\setminus\{i,j\}$ such that $i\dse_m j\cd C$; see \citet{ric02,sad11}.

This implies that two Markov equivalent maximal graphs must have the same skeleton, where the \emph{skeleton} of a graph results by replacing each edge by a full line.
\paragraph{Motivations behind using MAGs and its subclasses.} In this paper we aim to discuss Markov equivalence for the subclasses presented in Fig.\ \ref{fig:1110}. The conditions for internal Markov equivalence for RGs and SGs are very complex. However, in \citet{sad112} it was demonstrated how RGs can be mapped into a Markov equivalent SG, and how SGs can be mapped into a Markov equivalent AG. Notice that ADMGs are SGs without full lines, so by the same method one can map ADMGs into Markov equivalent AGs.

In addition, since Markov equivalent maximal graphs must have the same skeleton, conditions for Markov equivalence for MAGs are simpler than those for Markov equivalence for AGs. In \citet{ric02} it was shown how AGs can be mapped into a Markov equivalent MAG. Therefore, we map all types of stable independence graphs into MAGs and discuss the Markov equivalences for MAGs and its subclasses.

Notice that all subclasses of MAGs discussed here are maximal by nature. Therefore, for their Markov equivalence they must have the same skeleton.
\section{Internal and external Markov equivalences}
\subsection{Internal Markov equivalence for maximal ancestral graphs}
\begin{figure}[H]
            \centering
            \scalebox{0.5}{\includegraphics{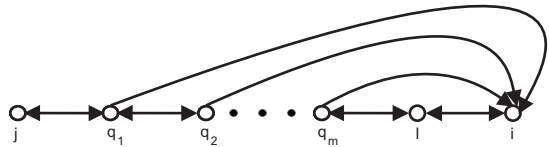}}
            \caption[]{\small{A discriminating path.}}
        \label{fig:4ex1}
\end{figure}

\begin{figure}[H]
\centering
\begin{tabular}{ccc}
\scalebox{0.5}{\includegraphics{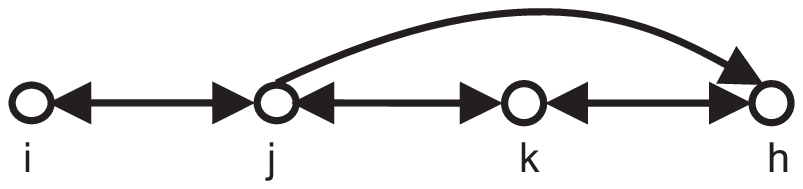}} &
\scalebox{0.5}{\includegraphics{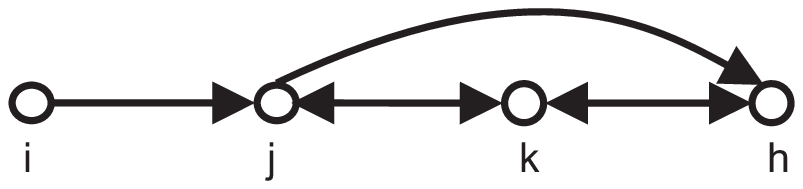}} &
\scalebox{0.5}{\includegraphics{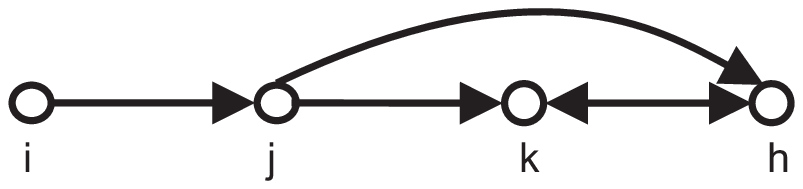}}\\
(a) & (b) & (c)
\end{tabular}
  \caption[]{\small{(a) A maximal ancestral graph $H_1$. (b) A maximal ancestral graph $H_2$ that is Markov equivalent to $H_1$.
  (c) A maximal ancestral graph $H_3$ that is not Markov equivalent to $H_1$ or $H_2$.}}
     \label{fig:4ex2}
\end{figure}

\begin{figure}[H]
            \centering
            \scalebox{0.5}{\includegraphics{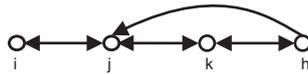}}
            \caption[]{\small{A non-minimal collider path $\langle i,j,k,h\rangle$ and a minimal collider path $\langle i,j,h\rangle$.}}
        \label{fig:4ex3}
\end{figure}

Thus far, there are two elegant results regarding Markov equivalence for MAGs available \citep{ali05,zha04}. These results use different definitions (colliders with order and minimal collider paths) and arguments. Even though it is not immediately obvious from their formulations, it can be shown that these are equivalent.
\paragraph{First result for Markov equivalence for MAGs.} In order to present the first theorem, we quote two definitions from \citet{ali05}. A path $\pi=\langle j,q_1, q_2,\dots, q_m,l,i\rangle$,
with $j$ not adjacent to $i$, is a \emph{discriminating path} for
$\langle q_m,l,i\rangle$ in $G$ if and only if, for
every node $q_n$, $1\leq n\leq m$ on $\pi$, i.e.\ excluding $j$, $i$, and $l$,
\begin{description}
\item[i)] $q_n$ is a collider on $\pi$; and
\item[ii)] $q_n\fra i$, hence forming a non-collider along the
path $\langle j, q_1,\dots,q_n,i\rangle$.
\end{description}
Fig.\ \ref{fig:4ex1} illustrates what a discriminating path looks like.\\
Let $\mathcal{D}_n$ be the set of \emph{triples of
order $n$} defined recursively as follows:
\begin{description}
\item[Order $0$:] A triple $\langle h,l,i\rangle\in\mathcal{D}_0$  if $h$ and $i$ are
not adjacent in $G$.
\item[Order $n+1$:] A triple $\langle h,l,i\rangle\in\mathcal{D}_{n+1}$ if
\begin{description}
\item[$1)$] $\langle h,l,i\rangle\notin\mathcal{D}_p$, for some $p < n + 1$ and
\item[$2)$] there exists a discriminating path $\pi=\langle j, q_1,\dots,q_m=h,l,i\rangle$ for $l$ in $G$, and each
of the colliders on the path: $\langle j,q_1,q_2,\rangle,\dots,\langle q_{m-1},q_m,l\rangle\in\bigcup_{p\leq n}\mathcal{D}_p$.
\end{description}
\end{description}
If $\langle h,l,i\rangle\in\mathcal{D}_n$ then the triple is said to have order $n$. A discriminating path is said to have order $n$ if every triple on the path has order at most $n$ and at least one
triple has order $n$. If a triple has order $n$ for some $n$
we then say that the triple has order, likewise for
discriminating paths.
\begin{theorem}\label{theorem:41}\citep{ali04}
MAGs $H_1$ and $H_2$
are Markov equivalent if and only if $H_1$ and $H_2$ have
the same skeleton and colliders with order.
\end{theorem}
In Fig.\ \ref{fig:4ex2} there are three MAGs with the same skeleton. In $H_1$ and $H_2$ since $i\nsim k$, the collider $\langle i,j,k\rangle$ is with order 0, whereas in $H_3$ this is not a collider. In $H_1$ and $H_2$, the collider $\langle j,k,h\rangle$ is with order 1. Therefore, we conclude that $H_1$ and $H_2$ are Markov equivalent, but these are not Markov equivalent to $H_3$.\\
\paragraph{Second result for Markov equivalence for MAGs.} In order to present the second theorem, we quote two definitions from \citet{zha04}. A path $\pi$ is called a \emph{collider path} if all its inner nodes are colliders on $\pi$. A collider path
$\pi=\langle i,B,j\rangle$ is called a \emph{minimal collider path} if $i\not\sim j$ and there is no proper subset $B'\subset B$ such that $\langle i,B',j\rangle$ is a collider path between
$i$ and $j$. If $i\sim j$ then we call $\pi$ a \emph{minimal collider cycle}. In the graph in Fig.\ \ref{fig:4ex3} the path $\langle i,j,k,h\rangle$ is a collider path, but it is not minimal collider since there exists the collider path $\langle i,j,h\rangle$, which is minimal collider.\\
\begin{theorem}\label{theorem:42}\citep{zha04}
MAGs $H_1$ and $H_2$ are Markov equivalent if and only if $H_1$ and $H_2$ have the same skeleton and minimal collider paths.
\end{theorem}
For the graphs in Fig.\ \ref{fig:4ex2}, by Theorem \ref{theorem:42} we can make the same conclusion as before. To observe this it is enough to check that $\langle i,j,k\rangle$ and $\langle i,j,k,h\rangle$ are the minimal collider paths of $H_1$ and $H_2$, whereas there is no minimal collider path in $H_3$. We, therefore, conclude that $H_1$ and $H_2$ are Markov equivalent, but these are not Markov equivalent to $H_3$.
\subsection{Internal Markov equivalence for subclasses of maximal ancestral graphs}
\begin{figure}[H]
\centering
\begin{tabular}{ccc}
\scalebox{0.5}{\includegraphics{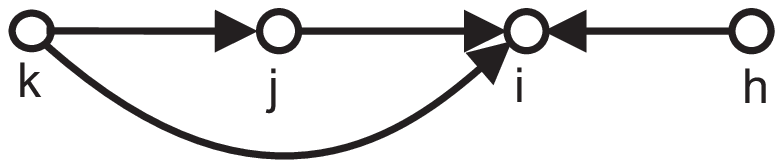}} &
\scalebox{0.5}{\includegraphics{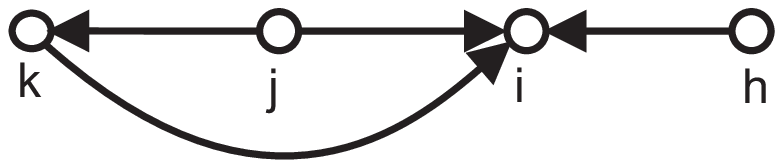}} &
\scalebox{0.5}{\includegraphics{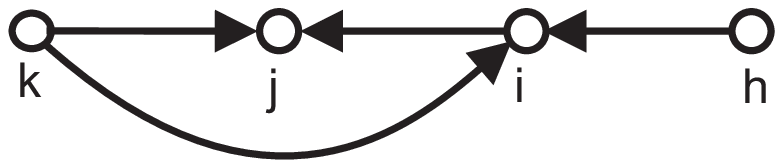}}\\
(a) & (b) & (c)
\end{tabular}
  \caption[]{\small{(a) A directed acyclic graph $G_1$. (b) A Markov equivalent directed acyclic graph $G_2$ to $G_1$.
  (c) A directed acyclic graph $G_3$ that is not Markov equivalent to $G_1$.}}
     \label{fig:41}
\end{figure}

\begin{figure}[H]
\centering
\begin{tabular}{ccc}
\scalebox{0.5}{\includegraphics{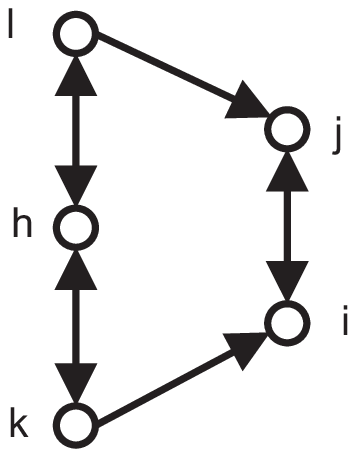}} &\nn
\nn\scalebox{0.5}{\includegraphics{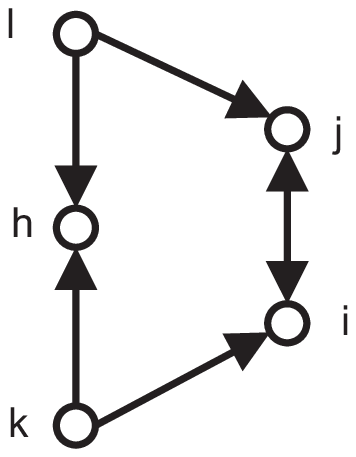}} &\nn
\nn\scalebox{0.5}{\includegraphics{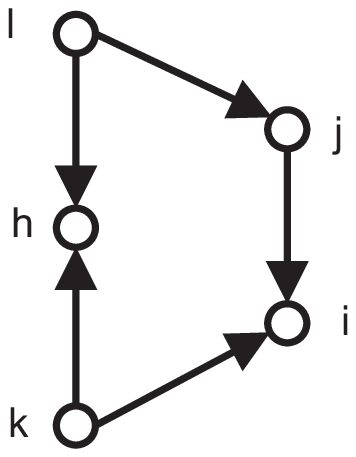}}\\
(a) & (b) & (c)
\end{tabular}
  \caption[]{\small{(a) A regression chain graph $H_1$. (b) A Markov equivalent regression chain graph $H_2$ to $H_1$.
  (c) An RCG (DAG) $H_3$ that is not Markov equivalent to $H_1$.}}
     \label{fig:4}
\end{figure}

\paragraph{Markov equivalence for DAGs.} First of all we recall a well-known result regarding Markov equivalence for DAGs.
\begin{prop}\label{prop:40}\citep{ver90,fry90}
DAGs $G_1$ and $G_2$ are Markov equivalent if and only if they have the same skeleton and unshielded collision V-configurations.
\end{prop}
In the example in Fig.\ \ref{fig:41} all three graphs have the same skeleton. In $G_1$ there are two unshielded collision V-configurations $\langle k,i,h\rangle$ and $\langle j,i,h\rangle$. In $G_2$ there are the same unshielded collision V-configurations. Therefore, these two graphs are Markov equivalent. The only unshielded collision V-configuration in $G_3$ is, however, $\langle k,i,h\rangle$. Hence this graph is not Markov equivalent to $G_1$ and $G_2$.\\
\paragraph{Markov equivalence for UGs and BGs.} The following proposition shows when two bidirected or undirected graphs are Markov equivalent.
\begin{prop}
Bidirected or undirected graphs $H_1$ and $H_2$ are Markov equivalent if and only if they are equal.
\end{prop}
\begin{proof}
In the case of the undirected graph, the result follows from Theorem \ref{theorem:41} and the fact that there is no collider in undirected graphs.

In the case of the bidirected graph, the result follows from Theorem \ref{theorem:42} and the fact that every path in bidirected graphs is a collider path.
\end{proof}
\paragraph{Markov equivalence for RCGs.} Since RCGs are a subclass of MAGs, we simplify the conditions for Markov equivalence for MAGs in order to obtain the conditions for Markov equivalence for RCGs.
\begin{prop}\label{prop:52}\citep{wers11}
RCGs $H_1$ and $H_2$ are Markov equivalent if and only if $H_1$ and $H_2$ have the same skeleton and unshielded collider V-configurations.
\end{prop}
\begin{proof}
We apply Theorem \ref{theorem:41} to RCGs and simplify its conditions in order to obtain the conditions of this theorem. The first condition of Theorem \ref{theorem:41}
 (having the same skeleton) is the same as the first condition of this theorem. Therefore, it is enough to prove that $H_1$ and $H_2$ have the same
 colliders with order if and only if they have the same unshielded collider V-configurations.

An unshielded collider V-configuration is by definition a collider with order. We prove that in RCGs a collider V-configuration that is a collider
with order is unshielded. This proves the proposition: Suppose that $\langle h,k,l\rangle$ is a collider with order and, for contradiction, is not
unshielded. By the definition of collider with order there exists a discriminating path $\langle x, q_1,\dots,q_p=h,k,l\rangle$
 for $k$. Hence $h\in\spo(k)$. In addition, if $l\in\pa(k)$ then $h\in\an(k)$, a contradiction by the definition of RCGs.
 Therefore, $l\in\spo(k)$, but again this is a contradiction since in RCGs for a collider V-configuration with two adjacent arcs
 $\langle h,k,l\rangle$, one endpoint ($h$) cannot be the parent of the other endpoint ($l$).
\end{proof}
In the example in Fig.\ \ref{fig:4} all three RCGs have the same skeleton. In $H_1$ there are three unshielded collider V-configurations $\langle l,h,k\rangle$, $\langle l,j,i\rangle$, and $\langle k,i,j\rangle$. In $H_2$ there are the same unshielded collider V-configurations. Therefore, these two graphs are Markov equivalent. The unshielded collider V-configurations in $H_3$ are, however, $\langle l,h,k\rangle$ and $\langle k,i,j\rangle$. Hence this graph is not Markov equivalent to $H_1$ or $H_2$.\\
\subsection{External Markov equivalence for subclasses of maximal ancestral graphs}
As a corollary of Proposition \ref{prop:52}, in order to check the external Markov equivalence for every two introduced subclasses of MAGs (excluding MAGs), i.e.\ RCGs, BGs, UGs, and DAGs, the conditions that are used for Markov equivalence for RCGs can be used. In some cases, if we suppose that the graphs satisfy the conditions for representational Markov equivalence, which will be introduced in the next section, then the conditions for external Markov equivalence can be simplified.
\begin{coro}
Every two of RCG, BG, UG, and DAG are Markov equivalent if and only if they have the same skeleton and unshielded collider V-configurations.
\end{coro}
\begin{proof}
The result follows from the fact that BGs, UGs, and DAGs are subclasses of RCGs.
\end{proof}
Notice that for Markov equivalence for a UG and a graph $H$ of another type the corollary states that there should be no collider V-configurations in $H$.
\section{Representational Markov equivalence and algorithms}
\paragraph{Structure of the section.} In this section, in each subsection, we deal with representational Markov equivalence for MAGs and its subclasses to a specific subclass. In each subsection we first introduce an algorithm for generating a graph of a specific subclass which is Markov equivalent to a given MAG. The algorithm is usually trivially simplified for subclasses of MAGs. We then introduce conditions for a MAG under which it is Markov equivalent to a graph from the given subclass. By simplifying these conditions we obtain the conditions for subclasses of MAGs under which they are Markov equivalent to a graph of the given subclass. Notice that representational Markov equivalence to the class of MAGs produces trivial results.
\subsection{Representational Markov equivalence to directed acyclic graphs}
\paragraph{Generating a DAG which is Markov equivalent to a given MAG.} We begin with an algorithm for generating a Markov equivalent DAG to a given MAG that satisfies the conditions of Lemma \ref{lem:5n1}.\begin{alg}\label{alg:530}
\text{(Generating a Markov equivalent DAG to a maximal ancestral graph $H$)}\\
Start from $H$.
\begin{enumerate}
  \item Apply the maximum cardinality search algorithm on $H[\full]$ to order the nodes.
  \item Orient the edges of $H[\full]$ from a lower number to a higher one.
  \item Replace unshielded collider V-configurations by unshielded collision V-configurations, i.e.\ replace $i\arc\circ\arc j$ and $i\arc\circ\fla j$ by
$i\fra\circ\fla j$ when $i\not\sim j$ .
	\item Order the nodes of the subgraph induced by arrows such that arrows are from higher numbers to lower ones.
	\item Order the nodes of the subgraph induced by arcs arbitrarily by numbers not used in the previous step if the number for the node does not already exist.
	\item Replace arcs by arrows from higher numbers to lower ones.
\end{enumerate}
Continually apply each step until it is not possible to apply the given step further before moving to the next step.
\end{alg}
\begin{lemma}\label{lem:5n1}
Let $H$ be a maximal ancestral graph. If $H[\full]$ is chordal and there is no minimal collider path or cycle of length $4$ in $H$ then Algorithm \ref{alg:530} generates a Markov equivalent DAG to $H$.
\end{lemma}
\begin{proof}
Denote the generated graph by $G$. Graph $G$ is directed since by Algorithm \ref{alg:530}, all edges are turned
into arrows. Since $H[\full]$ is chordal, the graph generated by
the perfect elimination order of the maximal cardinality search does not have a direction-preserving
cycle; see Section 2.4, \citet{bla92}.

In addition, the arrows present in the graph do not change by the algorithm. We show that there is no direction-preserving cycle after applying step 3: If, for contradiction, there is a shortest direction-preserving cycle after applying step 3 then a collider V-configuration $\langle j,k,i\rangle$ (say the $jk$-edge is an arc) should turn into a transition V-configuration after applying step 3. In this case there is an $hj$-edge for a node $h$ with an arrowhead pointing to $j$ and $h\not\sim k$. Since there is no minimal collider path or cycle of length $4$, $\langle j,k,i\rangle$ is shielded.  Notice that on the $ji$-edge there is an arrowhead pointing to $j$ since otherwise there is a minimal collider path or cycle of length $4$. This implies that the $ji$-edge is an arc, since otherwise a shorter direction-preserving path via the arrow from $i$ to $j$ is generated.

Since this edge should turn into an arrow from $j$ to $i$, $i\sim h$ and there is a node $l$ pointing to $i$ such that $l\not\sim j$.
Since $\langle h,j,i,l\rangle$ and $\langle k,j,i,l\rangle$ are collider paths (or cycles), on the $hi$-edge there is an arrowhead pointing to $i$, and $ki$ is an arc. To turn the $ki$-arc into an arrow from $i$ to $k$ there is a $kp$-edge with its arrowhead pointing to $k$, and $p\not\sim i$. Therefore, $\langle p,k,i,h\rangle$ is a minimal collider path (or cycle), a contradiction. Therefore, there is no direction-preserving cycle after applying step 3.

Therefore, the ordering of step 4 is permissible, and by step 6 there are obviously no direction-preserving cycles generated. Therefore, $G$ is acyclic.

Now we prove that $G$ is Markov equivalent to $H$: Since there is no minimal collider path of length 4 in $H$, by Theorem \ref{theorem:42}, $H$ is
Markov equivalent to $G$ if and only if they have the same skeleton and unshielded
collider V-configurations. Graph $G$ obviously has the same skeleton as that of $H$. In addition, an unshielded collider V-configuration in $G$ is an unshielded collider V-configuration in $H$. If, for contradiction, an unshielded collider V-configuration $\langle i,k,j\rangle$ in $H$ does not exist in $G$ then one of the arrowheads pointing to $k$, say on edge $ik$, must be removed by step 3. Therefore, there is an unshielded collider V-configuration $\langle l,i,k\rangle$ in $H$. Now $\langle l,i,k,j\rangle$ is a minimal collider path since $l\not\sim k$ and $i\not\sim j$, a contradiction.\end{proof}
It is easy to see that, for UGs, BGs, and RCGs that can be Markov equivalent to DAGs, the algorithm generates a Markov equivalent DAG to a given RCG, steps 1 and 2 of the algorithm generate a Markov equivalent DAG to a given UG, and steps 3-6 of the algorithm generate a Markov equivalent DAG to a given BG.

\begin{figure}[H]
\centering
\begin{tabular}{cc}
\scalebox{0.55}{\includegraphics{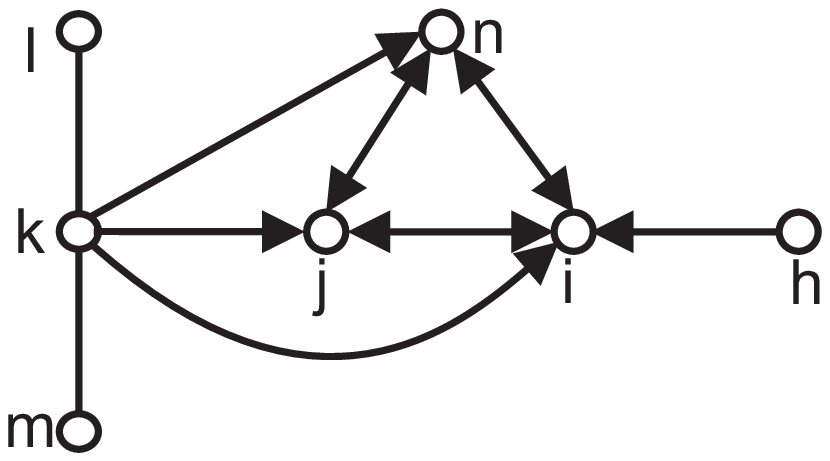}} &
\scalebox{0.55}{\includegraphics{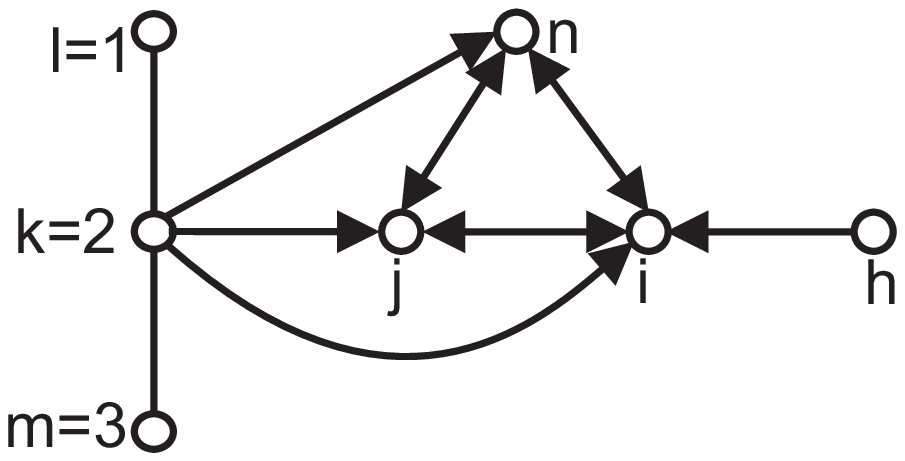}}\\
(a) & (b)\nl
\scalebox{0.55}{\includegraphics{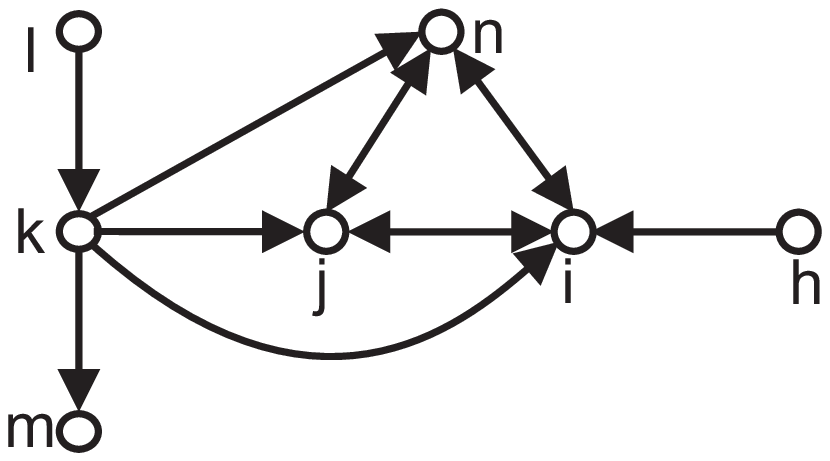}} &
\scalebox{0.55}{\includegraphics{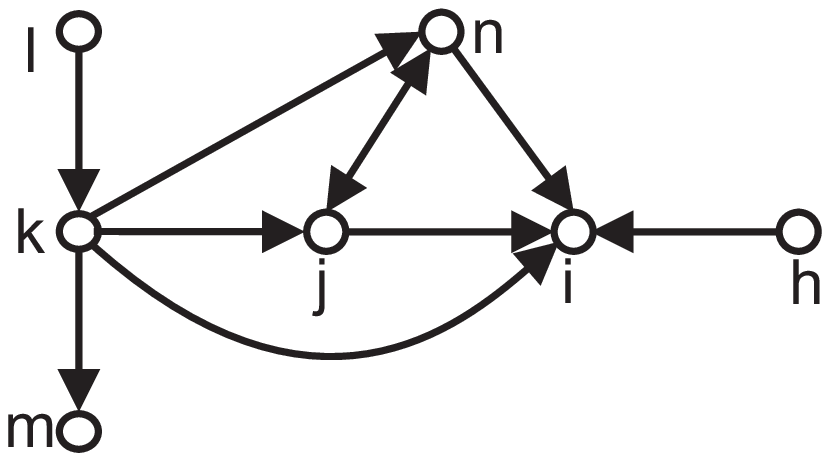}}\\
(c) & (d)\nl
\scalebox{0.55}{\includegraphics{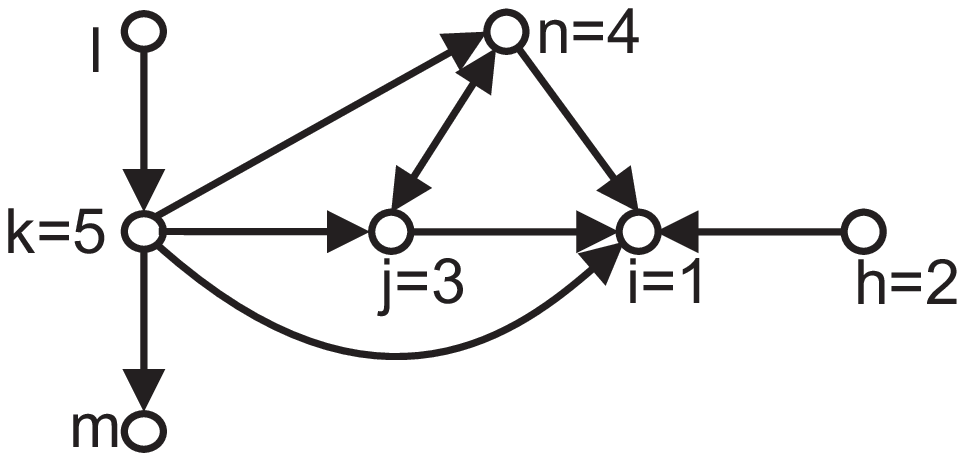}} &
\scalebox{0.55}{\includegraphics{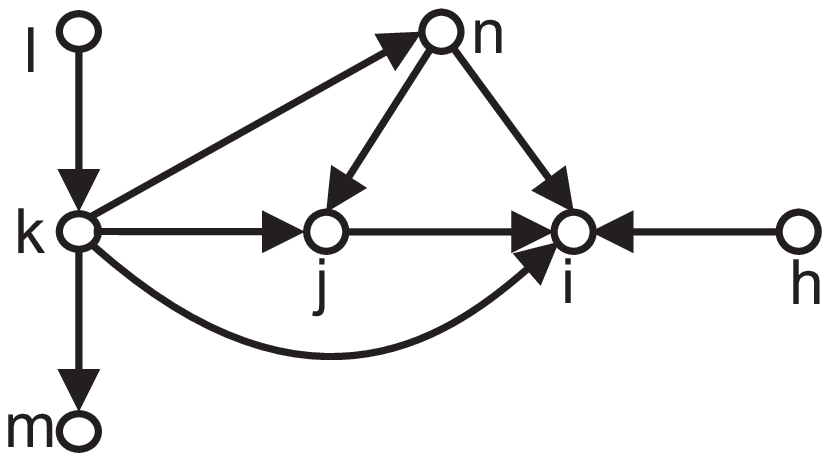}}\\
(e) & (f)
\end{tabular}
  \caption[]{\small{(a) A MAG. (b) The generated graph after applying step 1 of Algorithm \ref{alg:530}.
  (c) The generated graph after applying step 2. (d) The generated graph after applying step 3. (e) The generated graph after applying step 4.
  (f) The generated DAG after applying step 6.}}
     \label{fig:2ex2}
\end{figure}

Fig.\ \ref{fig:2ex2} illustrates how to apply Algorithm \ref{alg:530} step by step to a MAG.\\
\paragraph{Conditions for representational Markov equivalence for the class of MAGs and its subclasses to DAGs.} The following proposition shows that sufficient conditions for a given MAG, presented in Lemma \ref{lem:5n1}, are also necessary. The corollaries of this proposition illustrate the conditions under which RCGs, BGs, and UGs can be Markov equivalent to a DAG.
\begin{theorem}\label{prop:48}
A maximal ancestral graph $H$ is Markov equivalent to a DAG if and only if $H[\full]$ is chordal and there is no minimal collider path or cycle of length $4$ in $H$.
\end{theorem}
\begin{proof}
($\Rightarrow$) Suppose that the maximal ancestral graph $H$ is Markov equivalent to a directed acyclic graph $G$. By Theorem \ref{theorem:42} $G$ must have the same skeleton and minimal collider paths as those of $H$. Since there is no collider V-configuration in $H[\full]$, the corresponding induced subgraph of $G$ should have no unshielded collision nodes.
If, for contradiction, this subgraph contains an induced $C_n$, $n\geq 4$, then there exists a collision V-configuration on the cycle, otherwise there exists a direction-preserving cycle in $G$, which is not permissible. This collision V-configuration is unshielded since $n\geq 4$ and $C_n$ is chordless. This is a contradiction since $G$ and $H$ have the same skeleton. If $H$ contains
a minimal collider path or cycle $\pi$ then edges of $\pi$ cannot be oriented in $G$ to generate a collider path.

($\Leftarrow$) The result follows from Lemma \ref{lem:5n1}.
\end{proof}
We recall the following known statement for BGs as a corollary to the proposition.
\begin{coro}\label{coro:44}\citep{pea94}
A BG is Markov equivalent to a DAG if and only if it does not contain any $P_4$ or $C_4$ as induced subgraphs.
\end{coro}
\begin{proof}
For BGs, every path is a collider path, and every minimal collider path or cycle is a $P_4$ or $C_4$. Using these, the result follows.
\end{proof}
We also recall the following known statement for UGs as a corollary to the proposition; see \citet{lau96}.
\begin{coro}\label{coro:45}
A UG is Markov equivalent to a DAG if and only if it is chordal.
\end{coro}
\begin{proof}
For UGs, $H[\full]=H$, and there is no collider path in UGs. Using these, the result follows.
\end{proof}
The following corollary shows the conditions under which RCGs can be Markov equivalent to DAGs.
\begin{coro}\label{coro:46}
An RCG with chain component node sets $\tau_1,\dots,\tau_T$ is Markov equivalent to a DAG if and only if, $H[\full]$ is chordal and for $1\leq i\leq T$, the induced subgraph by
$\tau_i\cup\pa(\tau_i)$ does not contain any collider $P_4$ or $C_4$ as an induced subgraph.
\end{coro}
\begin{proof}
For RCGs, every collider path is in one of $\tau_i\cup\pa(\tau_i)$, $1\leq i\leq T$. In addition, in RCGs a minimal collider path or cycle is chordless. Using these, the result follows.
\end{proof}
\paragraph{Necessary conditions for representational Markov equivalence to DAGs.} Here we introduce some necessary conditions for Markov equivalence for MAGs, BGs, and RCGs to a DAG. For this purpose we use the following well-known graph theoretical result:
\begin{lemma}\label{lem:41}
If a graph $G$ contains no $P_4$ or $C_4$ as an induced subgraph then there is a node that is
adjacent to all other nodes.
\end{lemma}
\begin{coro}
Let $H$ be a MAG and $H[\arcc]=(V,E)$. If $H$ is Markov equivalent to a DAG then
there exists a node that is adjacent to all other nodes in $V\cup\pa(V)$.
\end{coro}
\begin{proof}
Graphs with no minimal collider paths or cycles of length $4$ do not contain a collider $P_4$ or $C_4$. In addition, every collider path in MAGs is in $V\cup\pa(V)$. Using these, the result follows from Theorem \ref{prop:48} and Lemma \ref{lem:41}.
\end{proof}
\begin{coro}
If a bidirected graph is Markov equivalent to a DAG then
there exists a node that is adjacent to all other nodes.
\end{coro}
\begin{proof}
The result follows from Corollary \ref{coro:44} and Lemma \ref{lem:41}.
\end{proof}
\begin{coro}
If an RCG is Markov equivalent to a DAG then in each $\tau_i\cup\pa(\tau_i)$, $1\leq i\leq T$,
there exists a node that is adjacent to all other nodes in $\tau_i\cup\pa(\tau_i)$.
\end{coro}
\begin{proof}
The result follows from Corollary \ref{coro:46} and Lemma \ref{lem:41}.
\end{proof}
\subsection{Representational Markov equivalence to undirected and bidirected graphs}
\paragraph{Generating a UG which is Markov equivalent to a given MAG.} By removing all arrowheads one generates a Markov equivalent UG to a given MAG that satisfies the condition of Lemma \ref{lem:5n2}.
\begin{lemma}\label{lem:5n2}
For a maximal ancestral graph $H$ with no unshielded collider V-configuration, removing arrowheads generates a Markov equivalent UG to $H$.
\end{lemma}
\begin{proof}
The generated graph is obviously a UG and is also the only UG that has the same skeleton as that of $H$. Neither $H$ nor the generated graph contains any minimal collider paths. This completes the proof.
\end{proof}
One can therefore observe that, for DAGs, BGs, and RCGs that can be Markov equivalent to UGs, removing arrowheads generates a Markov equivalent UG to the given graph.
\paragraph{Conditions for representational Markov equivalence for MAGs and its subclasses to UGs.} The following proposition shows that the sufficient condition for a given MAG, presented in Lemma \ref{lem:5n2}, is also necessary. The corollaries of this proposition illustrate the conditions under which BGs and DAGs can be Markov equivalent to a UG.
\begin{prop}
A maximal ancestral graph $H$ is Markov equivalent to a UG if and only if there is no unshielded collider V-configuration in $H$.
\end{prop}
\begin{proof}
($\Rightarrow$) Suppose that $H$ is Markov equivalent to an undirected graph $G$. Graphs $H$ and $G$ have the same skeleton and minimal collider paths, but $G$ has no minimal collider paths. Since an unshielded collider V-configuration is a minimal collider path, $H$ contains no unshielded collider V-configurations.

($\Leftarrow$) The result follows from Lemma \ref{lem:5n2}.
\end{proof}
One can also use this result for RCGs. Here we simplify the condition further for DAGs and BGs.
\begin{coro}
A directed acyclic graph $G$ is Markov equivalent to a UG if and only if there is no unshielded collision V-configuration in $H$.
\end{coro}
\begin{proof}
The result follows from the fact that the only type of colliders in DAGs is collisions.
\end{proof}
\begin{coro}
A BG is Markov equivalent to a UG if and only if it is complete.
\end{coro}
\begin{proof}
The result follows from the fact that all unshielded V-configurations in BGs are collider.
\end{proof}
\paragraph{Generating a BG which is Markov equivalent to a given MAG.} Replacing all edges by arcs generates a Markov equivalent BG to a given MAG that satisfies the condition of Lemma \ref{lem:5n3}.
\begin{lemma}\label{lem:5n3}
For a maximal ancestral graph $H$ with no unshielded non-collider V-configuration, replacing all edges by arcs generates a Markov equivalent BG to $H$.
\end{lemma}
\begin{proof}
The generated graph is obviously a BG and is also the only BG that has the same skeleton as that of $H$. All V-configurations in both $H$ and the generated graph are colliders. This completes the proof.
\end{proof}
 One can therefore observe that, for DAGs, UGs, and RCGs that can be Markov equivalent to BGs, replacing all edges by arcs generates a Markov equivalent BG to the given graph.
\paragraph{Conditions for representational Markov equivalence for MAGs and its subclasses to BGs.} The following proposition shows that the sufficient condition for a given MAG, presented in Lemma \ref{lem:5n3}, is also necessary.
\begin{prop}
A maximal ancestral graph $H$ is Markov equivalent to a BG if and only if there is no unshielded non-collider V-configuration in $H$.
\end{prop}
\begin{proof}
($\Rightarrow$) Suppose that $H$ is Markov equivalent to a bidirected graph $G$. Graphs $H$ and $G$ have the same skeleton and minimal collider paths, but every minimal path in $G$ is a minimal collider path. Since an unshielded non-collider V-configuration is a minimal but not a collider path, $H$ contains no unshielded non-collider V-configurations.

($\Leftarrow$) The result follows from Lemma \ref{lem:5n3}.
\end{proof}
One can also use this result for RCGs. Here we simplify the condition further for DAGs and UGs. The results have been known in the literature for long time, e.g., see \citet{pea94}.
\begin{coro}
A directed acyclic graph $G$ is Markov equivalent to a BG if and only if there is no unshielded non-collision V-configuration in $H$.
\end{coro}
\begin{proof}
The result follows from the fact that the only type of colliders in DAGs is collisions.
\end{proof}
\begin{coro}
A UG is Markov equivalent to a BG if and only if it is complete.
\end{coro}
\begin{proof}
The result follows from the fact that all unshielded V-configurations in UGs are non-collider.
\end{proof}
\subsection{Representational Markov equivalence to regression chain graphs}
\begin{figure}[H]
\centering
\begin{tabular}{cc}
\scalebox{0.55}{\includegraphics{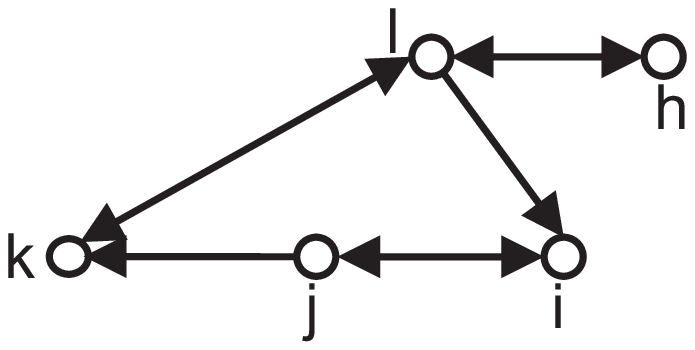}} &
\scalebox{0.55}{\includegraphics{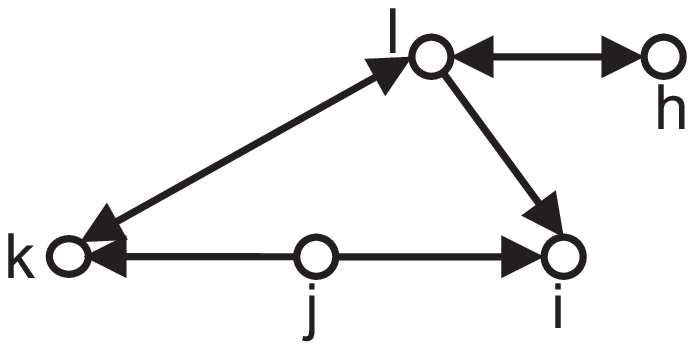}}\\
(a) & (b)
\end{tabular}
  \caption[]{\small{(a) A MAG. (b) The generated RCG after applying Algorithm \ref{alg:536}.}}
     \label{fig:2ex3}
\end{figure}

\paragraph{Generating an RCG which is Markov equivalent to a given MAG.} We begin with an algorithm for generating a Markov equivalent RCG to a given MAG that satisfies the conditions of Lemma \ref{lem:5n4}.
\begin{alg}\label{alg:536}
\text{(Generating a Markov equivalent RCG to a MAG $H$)}\\
Start from $H$.
\begin{enumerate}
  \item For a non-collider V-configuration $i\arc j\fra k$ on an arc-direction-preserving cycle, remove the arrowhead pointing to $j$ on the $ij$-edge when there is no unshielded collider V-configuration of form $\langle i,j,l\rangle$.
  \end{enumerate}
Continually apply this step until it is not possible to apply it further.
\end{alg}
\begin{lemma}\label{lem:5n4}
For a maximal ancestral graph $H$ with no arc-direction-preserving cycle on which every non-collider V-configuration $i\arc j\fra k$ is such that there is an unshielded collider V-configuration of form $\langle i,j,l\rangle$, Algorithm \ref{alg:536} generates a Markov equivalent RCG to $H$.
\end{lemma}
\begin{proof}
Denote the generated graph by $G$. To show $G$ is an RCG, it is enough to show that there is no arc-direction-preserving cycle in $G$. We know that the only arc-direction-preserving cycles in $H$ are those on which there is a non-collider V-configuration $i\arc j\fra k$ such that there is no unshielded collider V-configuration $\langle i,j,l\rangle$. In this case step 3 of the algorithm generates a source V-configuration, and therefore, destroys the arc-direction-preserving cycle.

We now prove that $G$ is Markov equivalent to $H$: First, we prove that minimal collider paths in $H$ remain unchanged in $G$. Suppose, for contradiction, that there is a minimal collider path $\pi$ of length $n$, $n\geq 3$, containing an $ij$-arc, and the arrowhead pointing to $j$ is removed by step 3 of the algorithm because of a V-configuration $i\arc j\fra k$ on an arc-direction-preserving cycle. Denote the three consecutive nodes on $\pi$ by $\langle i,j,l\rangle$. Since there is an arrowhead pointing to $j$ on the $jl$-edge, $i\sim l$. Since $\pi$ is minimal collider, there exists another node on $\pi$, say $h$ adjacent to $l$, $i$ is an endpoint of $\pi$, and the $li$-edge is an arrow from $l$ to $i$.

Now there is an arc-direction-preserving cycle $\langle i,j,l\rangle$ on which the only  non-collider V-configuration $j\arc l\fra i$ is such that there is a collider V-configuration  $\langle j,l,h\rangle$. Therefore, this collider V-configuration should be shielded, i.e.\ $h\sim j$. This edge is an arrow from $j$ to $h$ because $\pi$ is minimal collider. Therefore, $\langle i,j,l,h\rangle$ is a primitive inducing path, and since $H$ is maximal, $i\sim h$. This contradicts the fact that $\pi$ is a minimal collider path. Therefore, minimal collider paths in $H$ do not change by the algorithm.

In addition, a non-minimal collider path in $H$ cannot turn into a minimal collider path in $G$, since we know that in RCGs all collider paths are chordless.  This completes the proof.
\end{proof}
Fig.\ \ref{fig:2ex3} illustrates how to apply Algorithm \ref{alg:536} to a MAG.\\
\paragraph{Conditions for representational Markov equivalence for MAGs and its subclasses to RCGs.} Since UGs, BGs, and DAGs are subclasses of RCGs, there are no conditions needed under which they are able to be Markov equivalent to an RCG. The following proposition shows when a given MAG can be Markov equivalent to an RCG. The corollary of this proposition shows when UGs can be Markov equivalent to an RCG.
\begin{theorem}\label{prop:411}
A maximal ancestral graph $H$ is Markov equivalent to an RCG if and only if there is no arc-direction-preserving cycle on which every non-collider V-configuration $i\arc j\fra k$ is such that there is an unshielded collider V-configuration of form $\langle i,j,l\rangle$.
\end{theorem}
\begin{proof}
($\Rightarrow$) Suppose that $H$ is Markov equivalent to a multivariate regression chain graph $G$. Suppose, for contradiction, that there is an arc-direction-preserving cycle $\pi'$ on which every non-collider V-configuration $i\arc j\fra k$ is such that there is an unshielded collider V-configuration $\langle i,j,l\rangle$. If $\pi'$ has a chord $qr$ then there are two shorter cycles including the chord and nodes on $\pi'$ that are on one side of $q$ and $r$. One can observe that at least one of the two cycles has the same property as the property of $\pi'$ depending on whether there is an arrowhead pointing to $q$ or $r$ on $ij$-path. Hence, consider the minimal cycle $\pi$ in this sense, which is chordless.

On $\pi$ all collider V-configurations are unshielded. In addition, all collider V-configurations of form $\langle i,j,l\rangle$ are also unshielded. Therefore, since $H$ is Markov equivalent to $G$, in $G$ all these collider V-configurations should be preserved. Hence, all arcs on $\pi$ remain arcs in $G$. Moreover, by replacing the arrows on $\pi$ by arcs or by changing their directions a new unshielded collider V-configuration is generated. Therefore, arrows on $\pi$ are also unchanged in $G$. Therefore, $\pi$ exists in $G$. Since we know that arc-direction-preserving cycles are not permissible in RCGs, this is a contradiction.

($\Leftarrow$) The result follows from Lemma \ref{lem:5n4}.
\end{proof}
\section{Summary}
\paragraph{Summary of internal and external Markov equivalence for MAGs and its subclasses.} In Section 5, we showed that for internal and external Markov equivalences for subclasses of MAGs, excluding MAGs themselves, the conditions for Markov equivalences for DAGs can be generalised naturally by using colliders instead of collisions. In other words, two subclasses of MAGs are Markov equivalent if and only if they have the same skeleton and unshielded collider V-configurations.
\paragraph{Summary of representational Markov equivalence for MAGs and its subclasses.} The following table represents the summary of the conditions needed for representational Markov equivalence for MAGs and its subclasses. In addition, for each non-trivial case of table, we provided algorithms to generate a graph other types that is Markov equivalent to the graph of a given type.

The conditions presented in the table are for the graphs of the type indicated on the left column, which are to be Markov equivalent to a graph of the type indicated on the first row.
\begin{table}[H]\label{tab:51}
\caption{\small{Necessary and sufficient conditions on $H$ a graph of a subclass of maximal ancestral graph on the left column to be able to be Markov equivalent to a graph of the subclass of maximal ancestral graph on the top row.}}
\vspace{5mm}
{\small
\begin{tabular}{|c|c|c|c|c|c|}
  \hline
  $H\backslash $&  RCG & BG & UG & DAG\\\hline
   & No arc-dir-pr cycle with &   &  & $H[\full]$ Chordal;\\
  MAG & every $i\arc j\fra k$ s.t. &  No unshielded & No unshielded &no minimal collider\\
  & there is an unshielded&non-collider.&collider&path or cycle \\
  & collider $\langle i,j,l\rangle$&&&of length $4$\\\hline
  RCG & - & No unshielded  & No unshielded  & No collider $P_4$ or\\
  &  & non-collider & collider & $C_4$ in $\tau_i\cup\pa(\tau_i)$\\\hline
  BG & - & - & Complete & No $P_4$ or $C_4$
\\\hline
  UG & - & Complete & - & Chordal\\\hline
  DAG & - & No unshielded & No unshielded  & -\\
   &  & non-collision & collision & \\
  \hline
\end{tabular}}\\
\end{table}
\section*{Acknowledgments}
The author is grateful to Steffen Lauritzen and Nanny Wermuth for helpful comments.
\bibliographystyle{sjs}
\bibliography{bibber}
\vspace{10mm}
Kayvan Sadeghi, Department of Statistics, University of Oxford, 1 South Parks Road, Oxford, OX1 3TG, UK.
\nl
Email: sadeghi@stats.ox.ac.uk
\end{document}